\documentclass[twocolumn]{aastex62}

\shorttitle{The radio emission in J1302}

\def\ltsima{$\; \buildrel < \over \sim \;$}
\def\simlt{\lower.5ex\hbox{\ltsima}}
\def\gtsima{$\; \buildrel > \over \sim \;$}
\def\simgt{\lower.5ex\hbox{\gtsima}}
\newcommand{\msun}{{\rm\,M$_\odot$}}

\newcommand{\srcs}{{\rm\,J1302}}
\newcommand{\src}{{\rm\,J1302 }}

\begin{document}

\title{
Compact and variable radio emission from an active galaxy with supersoft X-ray emission 
}
\correspondingauthor{X. W.~Shu} 
\email{xwshu@ahnu.edu.cn}

\author{Lei Yang}
\affil{Department of Physics, Anhui Normal University, Wuhu, Anhui, 241002, China}
\author[0000-0002-7020-4290]{Xinwen~Shu}
\affil{Department of Physics, Anhui Normal University, Wuhu, Anhui, 241002, China}
\author{Fabao~Zhang }
\affil{Department of Physics, Anhui Normal University, Wuhu, Anhui, 241002, China}

\author[0000-0002-3698-3294]{Yogesh Chandola }
\affil{Purple Mountain Observatory, Chinese Academy of Sciences, No.10 Yuanhua Road, Qixia District, Nanjing 210023, China}

\author{Daizhong~Liu}
\affil{Max-Planck-Institut f\"ur Extraterrestrische Physik (MPE), Giessenbachstr. 1, D-85748 Garching, Germany
}

\author{Yi~Liu }
\affil{Purple Mountain Observatory, Chinese Academy of Sciences, No.10 Yuanhua Road, Qixia District, Nanjing 210023, China}

\author{Minfeng~Gu }
\affil{ Key Laboratory for Research in Galaxies and Cosmology, Shanghai Astronomical Observatory, Chinese Academy of Sciences, 80 Nandan Road, Shanghai 200030, China 
}

\author{Margherita Giustini }
\affil{  Centro de Astrobiologia (CAB), CSIC-INTA, Camino Bajo del Castillo s/n, Villanueva de la Ca\~nada, E-28692 Madrid, Spain}

\author{Ning Jiang}
\affil{Department of Astronomy, University of Science and Technology of China, Hefei, Anhui 230026, China}

\author{Ya-Ping~Li }
\affil{ Key Laboratory for Research in Galaxies and Cosmology, Shanghai Astronomical Observatory, Chinese Academy of Sciences, 80 Nandan Road, Shanghai 200030, China 
}

\author{Di~Li}
\affil{National Astronomical Observatories, Chinese Academy of Sciences, Beijing 100012, China}

\author{David~Elbaz }
\affil{Laboratoire AIM-Paris-Saclay, CEA/DRF/Irfu - CNRS - Universit\'e Paris Diderot, CEA-Saclay, 91191 Gif-sur-Yvette, France}

\author{Stephanie~Juneau }
\affil{NSF's National Optical-Infrared Astronomy Research Laboratory, 950 North Cherry Ave, Tucson, AZ 85719, USA}

\author{Maurilio~Pannella  }
\affil{Università di Trieste, Dipartimento di Fisica, Sezione di Astronomia, via Tiepolo 11, 34143 Trieste, Italy}

\author{Luming~Sun }
\affil{Department of Physics, Anhui Normal University, Wuhu, Anhui, 241002, China}

\author{Ningyu~Tang }
\affil{Department of Physics, Anhui Normal University, Wuhu, Anhui, 241002, China}

\author{Tinggui~Wang}
\affil{Department of Astronomy, University of Science and Technology of China, Hefei, Anhui 230026, China}

\author{Hongyan~Zhou}
\affil{Polar Research Institute of China, 451 Jinqiao Road, Shanghai, China}
\affil{Department of Astronomy, University of Science and Technology of China, Hefei, Anhui 230026, China}

\begin{abstract}
RX J1301.9+2747 is a unique active galaxy 
with supersoft X-ray spectrum that lacks significant emission at energies above 2 keV. 
In addition, it is one of few galaxies displaying quasi-periodic X-ray eruptions that recur on a timescale of 13-20 ks.
We present multi-epoch radio observations of RX J1301.9+2747 using GMRT, VLA and VLBA. 
The VLBA imaging at 1.6 GHz reveals a compact radio emission unresolved at a scale of $<$0.7 pc, 
with a brightness temperature of $T_{\rm b}$$>$$5\times10^{7}$ K. 
The radio emission is variable by more than a factor of 2.5 over a few days, 
based on the data taken from VLA monitoring campaigns.  
The short-term radio variability suggests that the radio emitting region has a size as small as $8\times10^{-4}$ pc, 
 resulting in an even higher brightness temperature of $T_{\rm b}$$\sim$$10^{12}$ K.
A similar limit on the source size can be obtained if the observed flux variability is not intrinsic and caused by 
the interstellar scintillation effect. 
The overall radio spectrum is steep with a time-averaged spectral index $\alpha=-0.78\pm0.03$ between 0.89 GHz and 14 GHz. 
These observational properties rule out a thermal or star-formation origin of the radio emission, and appear to be 
consistent with the scenario of episodic jet ejections driven by magnetohydrodynamic process. 
Simultaneous radio and X-ray monitoring observations down to a cadence of hours are required 
to test whether the compact and variable radio emission is correlated with the quasi-periodic X-ray eruptions. 


\end{abstract}

\keywords{Active galactic nuclei (16); Accretion (14); Radio jets (1347)}

\section{Introduction} \label{sec:intro}

Active galactic nuclei (AGNs) are powered by accretion of gas onto supermassive black holes (SMBHs) with 
$M_{\rm BH}\sim10^{6}-10^{9}$\msun. 
They are considered to be scaled-up versions of Galactic black hole X-ray binaries (XRBs, $M_{\rm BH}$ $\sim$10\msun), 
because of the similarities seen from the accretion flow, such as rapid X-ray variability \citep{Markowitz2005, McHardy2006}, 
the relation between the X-ray and radio emission \citep{Merloni2003, King2013}, and the correlation between 
the X-ray spectral index and Eddington ratio \citep{Yang2015, Ruan2019, Ai2020}. 
X-ray observations of XRBs have revealed the existence of two characteristic accretion states, 
namely soft and hard states \citep{Done2007}.  
The soft state is dominated by the thermal emission from an inner accretion disk. 
As accretion rate drops, the Comptonized coronal emission at higher energies comes to dominate the 
X-ray spectrum (hard state). 
It is well established that a flat-spectrum, compact jet is commonly observed during the hard state, which is however 
significantly quenched in the soft state \citep{Fender2004}. 
When sources transition from the hard to the soft accretion state, the radio emission begins to vary more dramatically, 
showing episodic jet ejection events  \citep{Miller-Jones2012, Bright2020}. 
Such spectral state transitions and the relationships between jet ejections and accretion
disk emission are still poorly understood in AGNs. 

 AGNs can be divided into radio-loud and radio-quiet populations according to the radio loudness parameter, which is defined as the ratio of the flux 
 densities between 6 cm and optical 4400\AA~($R\equiv f_{\rm 6~cm}/f_{4400\AA}$). 
 Radio-loud AGNs are conventionally classified as those with $R\simgt10$ \citep{Kellermann1989}. 
 It has been found that the radio loudness is anti-correlated with the Eddington ratios \citep{Ho2002, Greene2006}, albeit with a larger scatter, 
 implying that the jet radiation may be dependent on the accretion process. 
 This seems supported by the low radio-loud fraction \citep[$\sim$7\%,][]{Stepanian2003, Zhou2006} 
 observed in narrow-line Seyfert 1 galaxies (NLSy1s), a population of AGNs characterized by high accretion rates 
\citep{Collin2004, Komossa2008}.  
However, there are a few radio-loud NLSy1s exhibiting blazar properties \citep{Zhou2007, Yuan2008, Abdo2009, Foschini2020}, 
indicating that a relativistic jet can be formed at high accretion rate. 
This challenges canonical theories of jet formation that have been established in XRBs. 
Strictly speaking, the accretion states in NLSy1s are not analogous to the high-soft states seen in 
XRBs, as their X-ray spectra are still dominated by power-law emission originating from the corona.

The discovery of AGNs with supersoft X-ray emission in the past decade has drawn a considerable attention, 
including 2XMM J123103.2+110648 \citep{Terashima2012}, GSN 069 \citep{Miniutti2013}, and RX J1301.9+2747 \citep[hereafter J1302,][]{Sun2013}. 
Their X-ray spectra have weak or no hard X-ray emission at energies above $\sim$2 keV, and 
can be described by a dominant blackbody component with a temperature of $kT\sim0.05-0.1$ keV. 
This appears to be a close analog to the disk-dominated spectrum that
is typically seen in the high and soft states of XRBs \citep{Shu2017}. 
While such an extreme soft emission is unprecedented among AGNs or NLSy1s, they are more commonly seen in stellar 
tidal disruption event \citep[TDE;][]{Komossa2015, Saxton2020}. 
Indeed, two out of three known supersoft AGNs, 
 2XMM J123103.2+110648 and GSN 069, could be associated with TDEs \citep{Lin2017, Shu2018, Sheng2021}. 
Of particular interest, all supersoft AGNs are found to show regular X-ray variability 
in the form of quasi-periodic oscillations \citep[QPOs,][]{Lin2013, Song2020} or quasi-periodic eruptions \citep[QPEs,][]{Miniutti2019, Giustini2020}. 
The latter are characterized by short-lived flares of supersoft X-ray emission over a stable (quiescent) flux level, and recur on time scales of hours, 
representing a new cosmic phenomenon associated with extreme variability close to SMBHs, 
e.g., the radiation pressure instability in the accretion disk \citep{Pan2022}. 
This suggests a possible link on the part of the extreme variability phenomenon to the supersoft X-ray
component.


\src is the only supersoft AGN in which both QPOs and QPEs are detected \citep{Song2020, Giustini2020}. 
The QPO frequency is stable over almost two decades, suggesting that it may correspond to 
the high-frequency QPOs found in XRBs, adding further evidence of the similarity of \src to the XRBs. 
In this paper, we report the detection of a compact, variable radio emission in \srcs,  
possibly originating from optically-thin jet ejections. 
We adopt a cosmology of $\Omega_M$ = 0.3, $\Omega_{\lambda}$ = 0.7, and $H_0$ = 70 km s$^{-1}$ Mpc$^{-1}$ when computing luminosity distance.

\begin{deluxetable*}{cccccccc}
\tablewidth{0pt}
\tablehead{
\colhead{Project} & \colhead{Date} & \colhead{Array} & \colhead{$\nu${$_{\rm obs}$}} & \colhead{S$_{\rm int.}$} & \colhead{S$_{\rm peak}$} & \colhead{rms} & \colhead{Beam Size (PA)}\\
\colhead{} & \colhead{} & \colhead{} & \colhead{GHz} & \colhead{$\mu$Jy} & \colhead{$\mu$Jy/Beam} & \colhead{$\mu$Jy/Beam} & \colhead{arcsec × arcsec(deg)}}
\tablecaption{Summary of the radio observations of RX J1301.9+2747 \label{tab:table}}
\setlength{\tabcolsep}{3mm}{
\startdata
15A-349 & 05 July 2015 & VLA & 9.0 & 119.2  & 115.1 & 10.1 & 0.23 $\times$  0.18 (84.21)\\
 & 07 Aug 2015 & VLA & 9.0 & 154.7  & 154.4  & 10.4 & 0.18 $\times$ 0.17 (-22.89)\\
 & 07 Aug 2015 & VLA & 9.0 & 123.3  & 126.4  & 10.3 & 0.18 $\times$ 0.17 (-62.03)\\
 & 07 Aug 2015 & VLA & 9.0 & 141.6  & 143.4  & 9.8 & 0.19 $\times$ 0.18 (-87.24)\\
 & 08 Aug 2015 & VLA & 9.0 & 78.4  & 57.9  & 11.1 & 0.39 $\times$ 0.17 (68.58)\\
 & 09 Aug 2015 & VLA & 9.0 & 89.0  & 84.6  & 10.7 & 0.18 $\times$ 0.17 (-23.63)\\
 & 10 Aug 2015 & VLA & 9.0 & 90.4  & 98.2  & 10.0 & 0.18 $\times$ 0.18 (-65.56)\\
 & 10 Aug 2015 & VLA & 9.0 & 144.5  & 131.2  & 10.4 & 0.20 $\times$ 0.18 (88.07)\\
17B-027 & 04 Sep 2017 & VLA & 6.0 & 295  & 304 & 10.2 & 2.43 $\times$ 0.84 (-63.19)\\
18B-115 & 05 Jan 2019 & VLA & 14.0 & 183.7  & 174.8  & 9.2 & 2.12 $\times$ 1.19 (-69.92)\\
 & 06 Jan 2019 & VLA & 14.0 & 162.6  & 173.6  & 11.5 & 2.09 $\times$ 1.19 (-67.75)\\
 & 07 Jan 2019 & VLA & 14.0 & 188.5  & 154.8  & 9.0 & 2.16 $\times$ 1.22 (-68.15)\\
 & 08 Jan 2019 & VLA & 14.0 & 176.5  & 180.6  & 9.3 & 2.33 $\times$ 1.21 (-65.13)\\
 & 09 Jan 2019 & VLA & 14.0 & 183.4  & 158.9  & 8.4 & 2.00 $\times$ 1.24 (-69.84)\\
 & 10 Jan 2019 & VLA & 14.0 & 147.7  & 116.7  & 8.7 & 2.14 $\times$ 1.25 (-67.12)\\
 & 12 Jan 2019 & VLA & 14.0 & 179  & 139.8  & 9.2 & 2.08 $\times$ 1.24 (-67.62)\\
28$\_$039 & 31 May 2015 & GMRT & 1.4 & 764  & 735  & 23.2 & 2.40 $\times$ 1.97 (92.7)\\
BS255 & 14 Feb 2017 & VLBA & 1.6 & 670  & 559  & 32.5 & 0.012 $\times$ 0.0066 (25.22)\\
AM868 & 18 June 2006 & VLA & 1.4 & 866  & 779 & 117.2 & $\sbond$$^{\dag}$ \\
 VLASS$^{\ddag}$ & 25 Nov 2017 & VLA & 3.0 & 894  & 513  & 107 & 2.60 $\times$ 2.16 (-54.25) \\
AS110 & 17 Oct 2020 & ASKAP & 0.89 & 1880  & 1896  & 348.2 & 25.0 $\times$ 25.0 (0.00)\\
\enddata}
\tablecomments{
$^{\dag}$Due to the poor imaging quality, the beam size cannot be measured using CASA. 
$^{\ddag}$VLA Sky Survey (VLASS) consists of three-epoch observations, each separated by approximately a period of 32 months \citep{Lacy2020}. 
Only the epoch I data are used in this paper.  
The last three rows represent the archival VLA and Australian Square Kilometre Array Pathfinder (ASKAP) data (Section 3.3). 
}
\end{deluxetable*}

\section{OBSERVATION AND DATA REDUCTION} \label{sec:style}
\subsection{VLA}
\src was not detected by Faint Images of the Radio Sky at
Twenty cm (FIRST) using Karl G. Jansky Very Large Array (VLA), 
with a 5$\sigma$ upper limit on the peak flux of 0.95 mJy/beam. It was serendipitously detected in the deeper 
VLA 1.4 GHz imaging of Coma cluster, with a peak flux density of $0.78\pm0.11$ mJy \citep{Miller2009}. 
 To further study the origin of the radio emission, we observed \src with the VLA in three different bands, 
C, X, and Ku, {centered at 6.0 GHz}, 9.0 GHz, and 14.0 GHz,  respectively. 
The C-band observation was performed in the B configuration on 2017 Sep 4 (project code: 17B-027), 
and the data were divided into 16 spectral windows with a total bandwidth of 2 GHz.  
The X-band observations were carried out eight times in the A configuration between 2015 July 5 and 
Aug 10 (project code: 15A-349). 
At the Ku-band, monitoring observations with a daily cadence were carried out in the C configuration 
between 2019 Jan 5 and Jan 12 (project code: 18B-115). 
All the observations were phase-calibrated using the calibrator J1310+3220, 
and 3C 286 was used for bandpass and flux density calibration.  
The total integration time is 30-35 min per band, with on-source time of 20 min at C-band, 
10 min at X-band, and 12 min at Ku-band, respectively. 

The data were reduced using the Common Astronomy Software Applications (CASA, version 5.3.0) 
and the standard VLA data reduction pipeline (version 5.3.1). 
For the reduced data product, we inspected each spectral window and 
manually flagged channels affected by radio frequency interference (RFI).
The calibrated data were imaged using the {\tt CLEAN} algorithm with 
Briggs weighting and ROBUST parameter of 0, which helps to reduce 
side-lobes and achieve a good sensitivity. 
The final cleaned maps have a typical synthesized beam of 2\farcs4$\times$0\farcs8, 0\farcs2$\times$0\farcs2, 2\farcs1$\times$1\farcs2 and an rms noise of 10, $\sim$10-11, $\sim$8-12 $\mu$Jy/beam at 
C, X, and Ku-band, respectively. 
\src was clearly detected in all observations. 
We used the {\tt IMFIT} task in CASA to fit the radio emission component with a two-dimensional 
elliptical Gaussian model to determine the position, integrated and peak flux density. 
The radio emission at the three bands is unresolved and no extended emission is detected. 
The compactness of the radio emission is confirmed by the ratios of integrated and peak flux density, 
which are in the range 0.92--1.35, with a median of 1.09. 
For consistency, only peak flux densities are used in our following analysis.  
The VLA observation log and flux density measurements are presented in Table 1.


\subsection{GMRT}

J1302 was observed with the Giant Metrewave Radio Telescope (GMRT) at band 5 
(central frequency of 1.37 GHz) on 2015 May 31  (project code: 28$\_$039).
The GMRT band 5 data were divided into 512 channels across a bandwidth of 33.3 MHz. 
%
Flux calibration was conducted using 3C 147 and 3C 286, whereas the nearby source 3C 286 was also used to determine the 
complex gain solutions. 3C 147 was observed for $\sim$30 minutes in the beginning while 3C 286 was observed for  4 minutes 
after every 55 minutes of the J1302 scan and $\sim$30 minutes at the end of observation.  3C 147, 3C 286, and J1302 were observed for $\sim$30 min, $\sim$65 min and $\sim$ 9.2 h, respectively.
The data from the GMRT observations were reduced using CASA (version 5.6.1) following standard procedures and by using a pipeline adapted from the CAsa Pipeline-cum-Toolkit for Upgraded Giant Metrewave Radio Telescope data REduction \citep[CAPTURE;][]{Kale2021}. We began our reduction by flagging known bad channels, and the remaining RFI was flagged with the {\tt flagdata} task using the clip and tfcrop modes. In total, we flagged $\sim$20$\%$ of the data. We ran the task {\tt tclean} 
with the options of the {MS-MFS \citep[multi-scale multi-frequency synthesis,][]{Rau 2011} deconvolver,} two Taylor terms (nterms=2), and W-Projection \citep{Cornwell 2008} to accurately model the wide bandwidth and the non-coplanar field of view of GMRT. We also used a robust parameter of 0, imsize of 2500 pixels and cellsize of 0.5$\arcsec$ in task {\tt tclean}. 
In addition, we performed a few rounds of phase-only self-calibration to improve the fidelity of imaging. 
The final image has a synthesized beam of 2\farcs4$\times$2\farcs0. The GMRT flux density measurements are shown in Table 1.

\begin{figure*}[htb]
\plotone{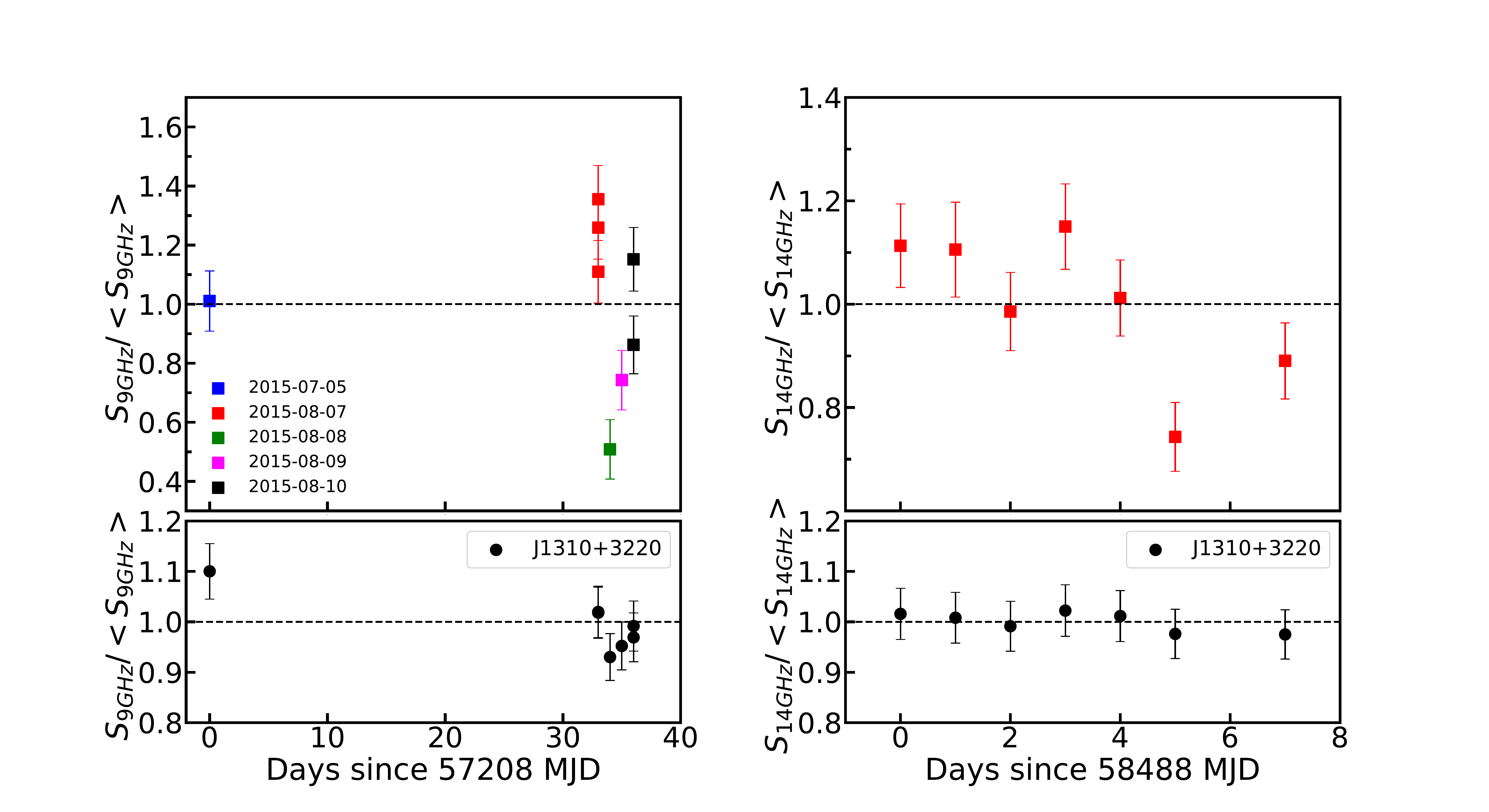}
\caption{{\it Left panel:} The normalized VLA peak flux densities relative to the mean flux at 9.0 GHz ($\left \langle S_{\rm 9GHz}\right \rangle$). 
The lower panel shows the same normalized peak flux densities but for the phase calibrator J1310+3220. 
{\it Right panel:} The same as left, but for the radio light curve observed at 14 GHz. 
\label{fig:f1}}
\end{figure*}

\subsection{VLBA}
We observed J1302 on 2017 Feb 14 using the Very Long Baseline Array (VLBA) with its 10 antennas (project code: BS255). 
The observing frequency was centered at 1.576 GHz in the L band. The observation was 
performed in the phase-referencing mode to a nearby strong compact radio source (J1300+28). 
Phase-reference cycle times were 4.5 minutes, with 3 minutes on-source and 1.5 minutes for the phase calibrator. 
We also inserted several scans of the bright radio source 3C 273 for fringe and bandpass calibration with an integration time of 2 minutes 
for each scan. 
The resulting total on-source time was 6 hours. 
To achieve sufficiently high imaging sensitivity, we adopted the observational mode RDBE/DDC to 
use the largest recording rate of 2 Gbps, corresponding to a recording bandwidth of 256 MHz in each of the dual circular polarizations. 
The data from the VLBA experiment were correlated with the DiFX software correlator \citep{Deller2011}. 
We used the NRAO AIPS software to calibrate the amplitudes and phases of the visibility data, following the standard procedure 
from the AIPS Cookbook\footnote{\url{http://www.aips.nrao.edu/cook.html}}. 
The calibrated data were imported into the Caltech DIFMAP package \citep{Shepherd1997} for imaging and model-fitting. 
The results are given in Table 1. 


\section{Results}
\subsection{Radio variability analysis}

Using the flux density between 6 cm and optical 4400 ${\AA}$ \citep{Shu2017}, we derived the radio-loudness parameter R $\sim$ 3.3. 
This suggests that J1302 is formally radio quiet, because a radio-loud object is usually defined to have R $>$ 10 \citep{Kellermann1989}. 
 We investigated the radio variability based on the radio flux measurements from the VLA monitoring observations at X and Ku-band. 
As shown in Table 1, 
{the radio emission of J1302 is variable on timescales of days to month (as short as a few hours within one day). }
Specifically, the maximum flux density is higher than minimum one by a factor of 2.6 at X-band, 
  and the variability amplitude is a factor of 1.5 at Ku-band. 
To further inspect the radio variability in individual observations, we plot in Figure \ref{fig:f1} the radio flux densities normalized to their time-average value for X-band (left) and Ku-band (right), respectively.  
We note that the error on the peak flux density given by {\tt IMFIT} task in CASA is likely underestimated, as it is smaller than the image off-source root-mean-square (rms).  
To be conservative, the flux errors shown in Figure \ref{fig:f1} were calculated as the sum in quadrature of map rms and calibration uncertainty that is assumed to 
 be of 5\% of the flux density \citep[e.g.,][]{Panessa2022}. 
It can be seen that the variability amplitude in the percent change in flux relative to the averaged one 
can be as high as $\sim$50\% at X-band, while it is $\sim$30\% at Ku-band. 
 In the lower panel, we also show the percent flux changes for the phase calibrator (J1310+3220), 
 which are at a level of only $\simlt$10\%. 
 This flux variability does not significantly correlated with that of \srcs, indicating that 
 the observed variability can not be dominated by the variability of phase or flux calibrators.

  To quantify the radio variability pattern of \src we calculated the debiased variability index $V_{\rm rms}$ \citep[e.g.,][]{Barvainis2005}. 
 $V_{\rm rms}$ is similar to the fractional variability $F_{\rm var}$ that is commonly used in analyzing X-ray light curves \citep[e.g.,][]{Vaughan2003}. 
 $V_{\rm rms}$ quantifies the variability amplitude in excess 
 of uncertainties as a percentage of the mean flux, 
 \begin{equation}
 V_{\rm rms}=\sqrt{\frac{S^{2}-\left \langle \sigma^{2}\right \rangle}{\left \langle  
 F_{\nu}\right \rangle^{2}}}
 \end{equation}
 where $S^{2}$ is the variance of the light curve, $\left \langle \sigma^{2}\right \rangle$ is the 
 mean squared flux errors which are the sum in quadrature of map rms and 5\% flux calibration uncertainty, 
 and $\left \langle F_{\nu}\right \rangle$ is the mean flux density.  
 A source is considered to be variable when $V_{\rm rms}>0$ by a significant amount. 
 For \src we measured $V_{\rm rms}\rm (9 GHz) = 26\pm$4 per cent and $V_{\rm rms}\rm (14 GHz) = 12\pm$3 per cent, respectively (Table 2). 

In addition, we performed $\chi^{2}$ analysis to test the significance of variability in the light curve differing from a constant. 
For a non-variable source, the value of reduced $\chi^{2}$ (i.e., $\chi^{2}$/d.o.f) is expected to be 1. 
For a given $\chi^{2}$ value, we can derive the variability significance $P_{\rm var}=1-$P($>$$\chi^{2}; \rm d.o.f)$, 
where $P$($>$$\chi^{2}; \rm d.o.f)$ is the probability to observe a $\chi^{2}$ larger than expected under the null hypothesis of 
no variability. 
According to the $\chi^{2}$ test, the X-band and Ku-band light curves shown in Figure \ref{fig:f1} 
 are variable at a confidence {level of $>$99.99\% and $>$99.94\%}, respectively, 
providing further evidence for the variability of \srcs. 
Note that \src was observed three times on 2015 Aug 7 at 9 GHz, separated by $\sim$1-2 hours, 
but the intraday variability amplitude was marginal with $V_{\rm rms}$(9 GHz)= 4.8$\pm$8.4 per cent. 
However, a larger intraday variability amplitude of $V_{\rm rms}$(9 GHz)= 20$\pm$7 per cent was found for the observations performed between 2015 Aug 9 and 2015 Aug 10, 
indicating that the radio emission can vary on a timescale as short as a few hours.  

 \begin{deluxetable}{ccccccc}
\tablewidth{0pt}
\tablehead{
\colhead{Peak flux} & \colhead{Mean} & \colhead{Std. dev.} & \colhead{$\chi^2$/d.o.f} & \colhead{$P_{\rm var}$} & \colhead{$V_{\rm rms}$ ($\%$)}}
\tablecaption{Variability analysis of the radio light curves. \label{tab:table}}
\setlength{\tabcolsep}{0.4mm}{
\startdata
\src $S_{\rm 9 GHz}$ ($\mu$Jy) & 113.9 & 32 & 51/7 & $>99.99$\% &26$\pm$4\\
\src $S_{\rm 14 GHz}$ ($\mu$Jy) & 157 & 23 &  24/6&  $>99.94$\%& 12$\pm$3\\
Calibrator $S_{\rm 9 GHz}$ (Jy) & 1.6 & 0.08 & 7/7 & 0.57 & 1.6$\pm$4.3 \\
Calibrator $S_{\rm 14 GHz}$(Jy) & 1.38 & 0.03 & 0.9/6 & 0.01 &  $\sbond$  \\
\enddata}
\tablecomments{Calibrator refers to J1310+3220. }
\end{deluxetable}

 \begin{figure*}[t!]
  \begin{center}
\includegraphics[width=0.75\textwidth]{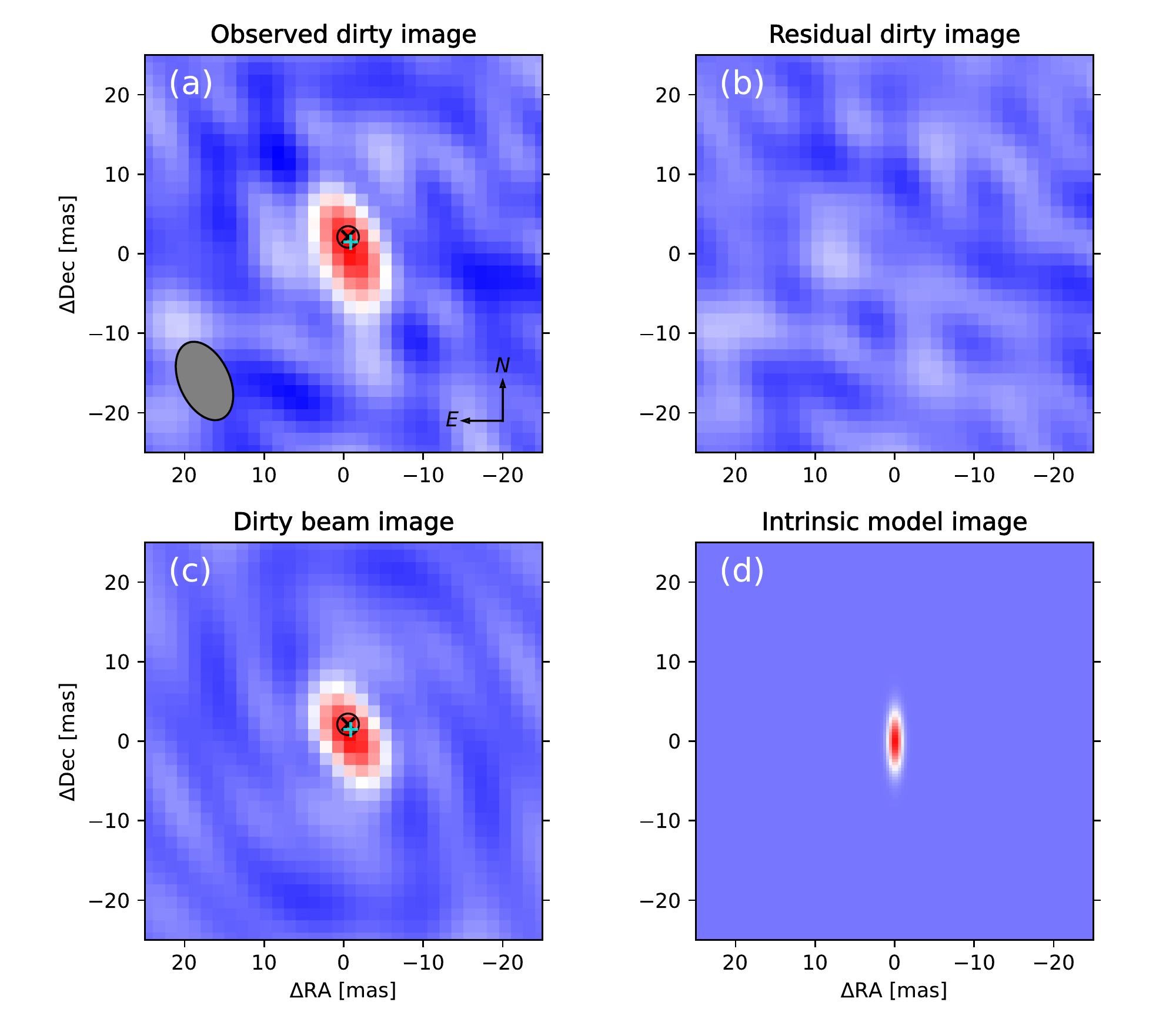}
 \hspace{0.1cm}
 \end{center}
  \vspace{-0.5cm}
\caption{VLBA 1.6 GHz image of \src and uv-plane source fitting result.  
Panel (a) and (c) shows the dirty source and beam images, respectively.  
The shape of the beam is shown in the left corner (gray filled ellipse) in panel (a). 
Panel (b) shows the residual map which was imaged from visibilities (subtracted a two-dimensional Gaussian model in the uv-plane). 
There are no additional emission components in the residual map shown. 
Panel (d) is the modeled intrinsic source image.  
All panels have the same color scale. 
The radio source position (cyan plus sign) detected by VLBA is centered at 
RA(J2000) = $13^{\rm h}02^{\rm m}00\fs13804$ $(\pm0\fs00025)$ and DEC(J2000) = +27$\arcdeg$46$\arcmin$57\farcs8489 ($\pm0\farcs00032$). 
The black cross marks the optical centroid obtained from {\it Gaia} DR3. 
The circle denotes the {\it Gaia} 1$\sigma$ positional error,  which is the sum in quadrature of the astrometric error and the uncertainty from astrometric excess noise. 
\label{fig:f2}}
 \vspace{0.2cm}
\end{figure*}

\subsection{Parsec-scale radio morphology}

In Figure \ref{fig:f2} (a), we show the VLBA image of \src at 1.6 GHz. 
The source appears to be unresolved 
which is very similar to the beam shape in size. 
Other than the central peak, no extended emission was seen above the image sensitivity of 0.098 mJy beam$^{-1}$ ($3\sigma$). 
 To measure the source size more accurately, 
we performed source fitting in the uv-plane to the total polarization intensity data using the UV\_FIT task in the 
GILDAS software package.  
We found no additional components in the residual map (Figure 2 (b)). 
The integrated flux density for the source in the uv-plane is $\sim$ 655 $\mu$Jy, in good agreement with the flux density measured 
in the image plane using the DIFMAP package (Table 1).   
As shown in Figure 2 (d), the VLBA source has a deconvolved size of 5.61 mas $\times$ 1.43 mas, indicating the extreme compactness 
of radio emission ($<$0.7 pc at the redshift of \srcs, $z=0.0237$).  
This is also consistent with the size measured using DIFMAP. 
The optical centroid reported by the {\it Gaia} DR3 \citep{Gaia2021} is marked as a black cross 
in Figure \ref{fig:f2}, which is RA(J2000)$=13^{\rm h}02^{\rm m}00\fs13806$, Dec(J2000)$=27\degr46\arcmin57\farcs
8496$, with $\sigma_{\rm ra}=0.18$~mas, $\sigma_{\rm dec}=0.14$~mas, and the astrometric excess noise of 1.35 mas\footnote
{Since \src is an extragalactic galaxy, we used the {\it Gaia} position without taking into account proper motions.}. 
Considering the positional uncertainties of {\it Gaia}, the radio peak does not show any positional offset with respect to the optical centroid, 
indicating that the emission components at the two bands arises probably from the same region very close to SMBH. 

The brightness temperature of compact radio emission can be estimated as \citep[e.g.,][]{Ulvestad2005}
\begin{equation}
T_b = \rm{1.8 \times 10^9 (1+z)\left (\frac{S_\nu}{1\,mJy} \right)\left(\frac{\nu}{1\,GHz} \right)^{-2}\left (\frac{{\theta_1}{\theta_2}}{1\,mas^2} \right)^{-1} K.}
\end{equation}
where $S_\nu$ is the peak flux density in mJy at the observing frequency $\nu$ in GHz, with $\theta_1$ and $\theta_2$ are the fitted FWHM of the major and minor axes of the Gaussian component in units of mas. Using the deconvolved size ($\theta$) derived from the VLBA image, we obtained a brightness temperature of $\sim$ 5.0 $\times$ $10^7$ K. 
Note that the pc-scale structure of \src is very compact and not resolved with the VLBA, only the upper limit on the source size 
can be constrained. 
Hence, the brightness temperature should be considered as a lower limit. 
Such a high brightness temperature allows us to exclude the star-formation origin for the radio emission since $T_b$ is typically 
lower than $10^6$ K as expected from star-forming processes \citep{Perez-Torres2021}. 
The emission from young supernova remnants can also be ruled out, as the 1.4 GHz radio luminosity of \src ($L_{\nu}=7.3\times10^{20}$ W Hz$^{-1}$) 
is higher than the majority of the radio supernova remnants, such as those studied in the starburst galaxy Arp 220 \citep{Varenius2019}. 
In fact, despite different resolutions, the radio flux densities measured with the VLBA and GMRT are comparable to that obtained with the VLA about a decade 
ago at the similar band \citep{Miller2009}, 
disfavoring the scenario of transients as the origin of the radio emission, such as a supernova explosion, gamma-ray burst or TDE. 

\subsection{Radio spectrum}
\begin{figure}[ht!]
\plotone{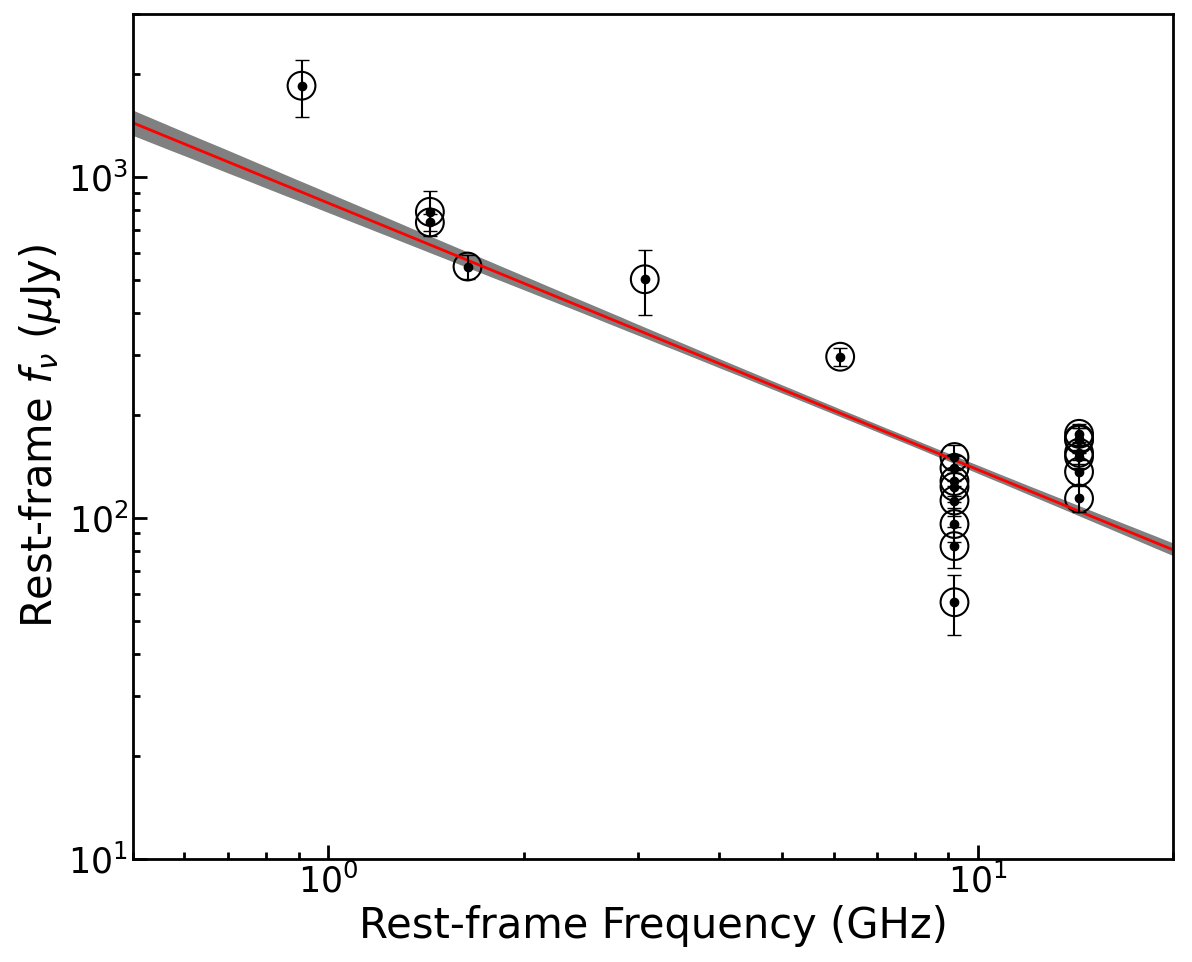}
\caption{The broad-band radio spectrum of \srcs. The flux density errors are calculated as the sum in quadrature of map rms and calibration uncertainty that is assumed to 
be 5\% of the flux density (Section 3.1). 
The red line shows the best-fit power-law spectrum using the data from all observations. 
The resulting radio spectral index is $\alpha= - 0.78 \pm$ 0.03.  
The error bars for the radio spectral index are estimated using Monte Carlo simulations (gray shaded region), 
assuming that the error on each flux density measurement follows a Gaussian distribution. 
\label{fig:f3}}
\end{figure}

Figure \ref{fig:f3} shows all the radio measurements available for \src based on our GMRT, VLA and VLBA projects. 
We also included the publicly available data from archival radio surveys such as Rapid ASKAP Continuum Survey  \citep[RACS,][] {McConnell2020}, 
VLA-Coma \citep{Miller2009}, and VLASS \citep{Lacy2020}.  
We retrieved the public radio maps and then measured the integrated and peak flux density with CASA, following the same procedures described in Section 2.1. 
The results are shown in Table 1. 
Although most observations were taken at different epochs and various resolutions, the radio flux density  
appear to decrease with frequency, indicating a steep radio spectrum between 0.89 GHz and 14 GHz. 
By fitting a power law spectrum ($S_{\nu} \propto \nu^{\alpha}$) to the data points, we obtained a spectral 
index $\alpha=- 0.78 \pm$ 0.03 (red line in Figure \ref{fig:f3}). 
This value is consistent with the typical radio spectral index of $\alpha=-0.6$ for the Palomar Seyfert galaxies \citep{Ho2001}. 
Note that using the data obtained from the quasi-simultaneous observations only, i.e., the GMRT and VLA observations performed on 2015 May-Aug, 
the radio spectrum of \src can still be described by a steep power law with slope $\alpha=- 1.03 \pm$ 0.04. 
Therefore, the emission from \src has a steep radio spectrum, even considering the time variability, 
suggesting that it may be related to optically thin synchrotron emission.  
We will discuss the implications of the steep radio spectrum in Section 4. 

\section{Discussion} 
We report radio variability on daily to month timescales in the supersoft AGN J1302 through VLA monitoring 
campaigns in 15 epochs on 2015 July-Aug and 2019 Jan. In addition, we detect an unresolved radio emission at a mas-scale with the VLBA, 
corresponding to $<$0.7 pc at the redshift of J1302. 
J1302 is one of few AGNs with X-ray QPEs, and has a low black hole mass ($M_{\rm BH}\simlt10^{6}$\msun) accreting at a 
high rate \citep[$L_{\rm bol}/L_{\rm Edd}\simgt$0.1,][]{Shu2017}. 
The radio light curve is variable with a rms amplitude of 26 $\pm$ 4 per cent and 12 $\pm$ 3 per cent with respect to the mean flux density
at X-band (9 GHz) and Ku-band (14 GHz), respectively. 
Daily flux variation up to a factor of $\sim$2.6 is also observed at X-band. 
Such highly variable radio emission is an unusual property for \src as a radio-quiet AGN (Section 3.1). 
Radio emission from radio-quiet AGNs could arise from different physical processes \citep{Panessa2019}, including 
star-formation, outflow, jet and accretion-disk corona. 
We will discuss these scenarios in detail according to the radio flux variability, morphology, size, brightness 
temperature, and time-averaged spectral index. 

\subsection{ Extrinsic variability caused by interstellar scintillation? }
The radio flux variability, if it is intrinsic to the AGN, would be useful to distinguish the 
sources of radio emission. 
We first consider whether the observed variability might be induced by the effect of refractive interstellar scintillation (ISS). 
This process occurs when radio waves propagate through an inhomogeneous plasma 
of our Galaxy, which could cause intraday variability in some AGNs with compact 
radio emission \citep{Lovell2003, Rickett2007}. The amount and timescale of radio 
variation caused by ISS depend on the Galactic electron column density along the line 
of sight and the observing frequency. 
Using the NE2001 free electron density model developed by \citet{Cordes2002}\footnote{https://pypi.org/project/pyne2001/}, 
and the Galactic dispersion measure (DM)\footnote{\src has a galactic coordinate of $(l, b=49.138781^{\circ} , 87.566837^{\circ})$. 
We used the DM of 9.25 cm$^{-3}$ pc for \srcs, which was derived from the pulsar J1239+2453 ($b>86^{\circ}$) in the ATNF Pulsar Catalogue.} along the line of sight to \srcs, 
we find that, at the position of \srcs, the transition frequency between strong and weak scattering regime 
is $\nu_{0}$$\sim$4.9 GHz. 
Hence \src is in the weak-scattering regime at 9 GHz and 14 GHz. 
In this case, we computed the fractional modulation index due to ISS \citep{Walker1998}\footnote{In \citet{Walker1998}, the modulation index $m$ is used to 
calculate the variability amplitude caused by ISS, which is defined as the ratio of the rms deviation to the mean value of the observed 
flux densities, 
$m=\sqrt{\frac{S^{2}}{\left \langle  F_{\nu}\right \rangle^{2}}}$. }  
which is {$m=0.42$ at 9 GHz and $m=0.23$ at 14 GHz}, respectively, comparable to that observed in \srcs. 
This amount of modulation is expected to occur on {a time-scale of $t_{\rm F}\sim2(\nu_{0}/\nu)^{1/2}$ h $\sim$1-2 h.} 
While \src is found to vary at 9 GHz on a timescale as short as a few hours, the variability amplitude 
can be as low as $V_{\rm rms}=4.8\pm8.4$ per cent, which seems inconsistent with the prediction of the ISS effect. 
On the other hand, if ISS is the main mechanism responsible for the flux density variation, 
the angular size of the radio source should be similar to the first Fresnel zone, {$\theta_{\rm F}\sim8/\sqrt{D\nu}\sim$ 2.4-3.0 $\mu$as}, 
where $D$ (in kpc) is taken as the NE2001 model distance for DM$=9.25$ cm$^{-3}$ pc, and $\nu$ is the observed frequency. 

\subsection{ Sub-kpc outflow}

Although the radio emission is variable and observed at different epochs and resolutions, we can obtain a time-averaged 
radio spectral index of $\alpha= - 0.78 \pm$ 0.03 between 0.89 GHz and 14 GHz (Figure \ref{fig:f3}), suggestive of 
an optically-thin steep spectrum. 
Note that if using the data from GMRT 1.4 GHz, {VLA 3 GHz, 6 GHz, and 14 GHz} observations which 
have similar resolutions (i.e., $\sim$2\arcsec), {a consistent radio steep spectrum can be {obtained ($\alpha= - 0.70 \pm$ 0.03)}.} 
\citet{Laor2019} suggested that a sub-kpc outflow interaction with the ambient interstellar medium 
could cause the optically-thin radio emission in radio-quiet AGNs. 
However, if such a nuclear outflow interaction were present, it would be resolved by our high-resolution VLBA observation  
into a number of clumpy structures or extended diffused emission \citep[e.g., ][]{Yang2021, Yao2021}. 
{It is obviously incompatible with the VLBA imaging of \srcs. 
On the other hand, there are no blue-shifted absorption or emission lines in the X-ray and optical spectra suggestive of outflows. 
Hence the outflow interpretation 
for the radio emission seems disfavored.} 

\subsection{ Steady and continuous jet}

Based on the high-resolution VLBA observation, \src has a high brightness temperature of $\simgt$5$\times$ $10^7$ K, 
which rules out the origin of radio emission from star-forming region or thermal processes (Section 3.2). 
Such a brightness temperature can place \src at the high end of the $T_{\rm b}$ range for the radio 
cores detected in local Seyfert galaxies \citep[$D\simlt22$ Mpc, ][]{Panessa2013}. 
If the flux variations in the radio light curve (Figure 1) are intrinsic to the AGN, 
the minimum variability on a timescale of one day can place a constraint on the source size of less than $8\times10^{-4}$ pc using the light-crossing time. 
Such a small source size suggests that the radio variability might originate from the innermost region of the AGN. 
Adopting the mean flux density at 9 GHz, this implies that the brightness temperature of the variable radio emission is 
$\simgt$9.4 $\times$ $10^{11}$ K. 
Therefore, the brightness temperature is comparable to the inverse Compton catastrophe limit of $\sim$ $10^{12}$ K \citep{Kellermann1969}, 
at which an emission region will radiate away most of its energy in X-rays via inverse Compton scattering in a timescale of days. 
The brightness temperature is however an order of magnitude higher than the equipartition temperature of 5$\times$ $10^{10}$ K \citep{Readhead1994},  
which is also usually used to indicate the intrinsic brightness temperature of the radio core in AGNs.  
 The value was estimated by assuming that there is equipartition of energy between the radiating particles and the magnetic field,
which is derived from a spectral cutoff due to synchrotron self-absorption \citep[see details in][]{Readhead1994}. 
Note that if the radio variability is extrinsic and caused by ISS, a similar limit on source size 
{($r_{\rm src}\sim1.2-1.4\times10^{-4}$ pc constrained by the first Fresnel zone)} hence the brightness temperature 
can be obtained. 
This implies that the radio emission could be associated with a jet.  
In this case, the jet could have a Doppler factor of $\sim$1--10, depending on which 
intrinsic brightness temperature is taken. 
Hence, the beaming effect is not significant. 

If the radio emission is from a steady and continuous jet, standard presumptions (by analog with XRBs) 
would be that \src is in the low-hard state which is dominated by a hard X-ray power-law component at energies above 2 keV \citep{Fender2004}. 
However, X-ray emissions from \src have been supersoft, with more than 90 per cent of photons 
having energies below 1 keV 
\citep{Sun2013, Shu2017, Giustini2020}, resembling the high/soft state observed in XRBs.  High-resolution VLBI observations have found evidence of the presence of a jet feature in many local Seyfert galaxies \citep{Panessa2013} 
as well as radio quiet NLSy1s with high accretion rates \citep[e.g.,][]{Doi2013, Yao2021}, either extending up to kpc scales or being confined to the inner pc regions 
due to the jet interaction with the circumnuclear gas medium.  
We do not find any extended structures in \src predicated by steady jets with the high-resolution VLBA observation. 
This suggests that the steady-jet model cannot account for the radio emission. 

\subsection{Episodic jet ejections}
Given its high accretion rate ($L_{\rm bol}/L_{\rm Edd}\simgt$0.1) and steep radio spectrum, one possible identity for \src is transitioning from 
the hard to soft state 
accompanied by an isolated radio ejection event, as observed in some XRBs \citep{Corbel2001, Fender2004, Bright2020}. 
It should be noted that \src is one of the few AGNs with X-ray QPEs detected, which are 
characterized by repetitive short-lived eruptions in the light curves. 
Strictly speaking, the X-ray eruptions in \src seem to be irregularly spaced in time rather than quasi-periodic \citep{Giustini2020}, 
which may be linked to the evolving corona as well as jet launching process \citep{Wilkins2015, Gallo2019}. 
Although no hard X-ray detected in \srcs, the ratio of 5 GHz to X-ray (0.3-2 keV) luminosity in the flare state is 
$3.4\times10^{-5}$, consistent with the relation for stars with active coronae \citep{Laor2008}. 
In the quiescent state, the luminosity ratio is an order of magnitude higher ($L_{\rm 5 GHz}/L_{\rm 0.3-2 keV}=7.7\times10^{-4}$), 
but still much lower than for radio-loud AGNs. 
This analog suggests that a magnetically heated corona may be responsible for the radio emission of \srcs. 
However, theoretical studies suggest that a flat synchrotron radio spectrum (between $\sim1-300$ GHz) 
would be expected from an X-ray corona \citep{Raginski2016}, which seems inconsistent with the steep spectrum 
observed in \src (Section 3.3).  
It should be noted that steep radio cores at mas scales have also detected in other Seyfert galaxies, though not being 
prevalent \citep[e.g.,][]{Panessa2013, Congiu2020}, which usually indicate the presence of optically thin synchrotron emission.  
 
 On the other hand, episodic ejections of magnetized plasmoids in the innermost region of the AGN  
 could be invoked to explain the radio emission. 
By analogy with the coronal mass ejection in solar physics \citep{Lin2000}, 
\citet{Yuan2009} proposed a magnetohydrodynamical model for episodic ejection of plasmoids from black
holes associated with the closed magnetic fields in an accretion flow.  
In the model, magnetic field loops are twisted to form flux ropes because of the turbulence of the accretion flow. 
With the magnetic energy gradually accumulated to reach a threshold, the flux rope loses its equilibrium and 
is thrust outwards in a catastrophic way, leading to magnetic reconnection.  
The magnetic energy is released in this process and converted into the energy of thermal electrons in plasmoids 
associated with the ejection of the flux rope, which then emit strong synchrotron radiation. 
Since the plasmoid ejecta can be considered as single blobs, they expand almost adiabatically
after leaving the accretion disk and can quickly become optically thin in the radio band \citep{Yuan2009}, 
explaining the steep spectrum.  
If this is the case, the episodic X-ray eruptions in \src may be regulated by magnetohydrodynamic process as well \citep{Li2017}. 
A high degree of polarization and polarization angle change would also be expected during the multi-band flares. 
Simultaneous radio and X-ray monitoring observations, particularly in the X-ray eruption state, are required 
to search for any correlated variability between the two bands, which is a hallmark of 
the coronal ejections as the origin of short-lived X-ray flares and will be presented elsewhere.

\subsection{ Comparison to other supersoft AGNs }

In addition to J1302, two other AGNs having supersoft X-ray spectra\footnote{Note that while two more galaxies are found to show QPEs with similar 
supersoft X-ray spectra \citep{Arcodia2021},  they are possibly not be associated with AGNs.} are 2XMM J123103.2+110648 and GSN069. 
The latter two sources were also observed with the VLA on 2017 Nov, in the same program as J1302 (17B-027). 
GSN 069 has weak radio emission, with a flux density of $S_{\rm 6GHz}=61\pm25$ $\mu$Jy, while J1231+1106 was 
not detected with a {5$\sigma$ upper limit of 38.6 $\mu$Jy}. 
Following the X-ray QPEs detected in GSN 069, an X-ray/radio campaign has been carried out, 
and no significant radio variability during QPEs is found \citep{Miniutti2019}. 
The inferred radio spectral index is steep ($\alpha=-$0.7), consistent with that measured in \srcs, suggesting 
a radio origin from optically thin synchrotron emission. 
However, the ratio of 5 GHz to X-ray (0.3-2 keV) luminosity in the QPE state 
is  $4.5\times10^{-7}$, about two orders of magnitude lower than that for \srcs, 
indicating either the radio emission or the flaring X-ray emission has different origins between the two objects. 
Owing to its faint radio flux density ($<$100 $\mu$Jy), using the VLBA to image the pc-scale structure of radio emission in 
GSN 069 is challenging. 


\section{Conclusion}
We present GMRT, VLA, and VLBA observations of the supersoft AGN J1302, 
which revealed compact radio emission with a size of $<$0.7 pc. 
The radio emission is variable on a daily timescale at both 9 GHz and 14 GHz, 
implying that the radio emitting region can have a size as small as $8\times10^{-4}$ pc, with a high 
brightness temperature of $T_{\rm b}$$\sim$$10^{12}$ K. 
If the observed flux variability is not intrinsic and caused by 
the interstellar scintillation effect, a similar limit on the source size can also be obtained ($\sim1.2-1.4\times10^{-4}$ pc). 
The radio spectrum is steep with a time-averaged spectral index $\alpha=-0.78\pm0.03$ between 0.89 and 14 GHz, 
suggesting that it may be related to optically thin synchrotron emission.
These observational properties rule out the radio origin from star-formation activities, sub-kpc outflows, or AGN corona. 
Optically thin jet ejection driven by magnetohydrodynamic process seems to be the most favored scenario.  
The magnetohydrodynamic model can be tested with future more sensitive polarimetric observations, 
which predicts a high degree of polarization in the radio emission. 
\acknowledgments{
We thank the anonymous referee for his/her helpful comments that improve the work. 
The data presented in this paper are based on observations
made with the Karl G. Jansky Very Large Array from the program VLA/15A-349, VLA/17B-027, and VLA/18B-115, 
the Very Long Baseline Array from the project VLBA/BS255, 
and the Giant Metrewave Radio Telescope from the project 28\_039. 
We thank the staff of the VLA, VLBA, and GMRT that made these observations possible. 
 The National Radio Astronomy Observatory is a facility of the
National Science Foundation operated under cooperative agreement
by Associated Universities, Inc.  
GMRT is run by the National Centre for Radio Astrophysics of the Tata Institute of Fundamental Research.
 The work is supported by National Science Foundation of China (NSFC) through grant No. 11822301, 12192220, 12192221, and 11833007. 
  M.F.G acknowledges support from the NSCS (grant No. 11873073), Shanghai Pilot Program for Basic Research Chinese
 Academy of Science, Shanghai Branch (JCYJ-SHFY-2021-013), and the science research grants from the China
Manned Space Project with NO. CMSCSST-2021-A06.  
Y.C. acknowledges the support from the NSFC under grant No. 12050410259, and Center for Astronomical Mega-Science, 
Chinese Academy of Sciences, for the FAST distinguished young researcher fellowship (19-FAST-02), and 
MOST through grant no. QNJ2021061003L. 
Y.L. acknowledges the support from Major Science and Technology Project of Qinghai Province (2019-ZJ-A10). 
M.G. is supported by the ``Programa de Atracci\'on de Talento'' of the Comunidad de Madrid, grant number 2018-T1/TIC-11733. 
}

\software{CASA \citep[v5.3.0 and v5.6.1; ][]{McMullin2007}, AIPS \citep{Greisen2003}, DiFX software correlator \citep{Deller2011}, DIFMAP \citep{Shepherd1997}, GILDAS\footnote{\url{https://www.iram.fr/IRAMFR/GILDAS}}
 }

\end{document}